\documentstyle[pra,aps,epsfig,twocolumn]{revtex}
\begin{document}
\draft
\title{Realistic Semiconductor Heterostructures design using Inverse Scattering}
\author{${}^{\dag }$Luca Perotti and Daniel Bessis}
\address{CTSPS, Clark-Atlanta University, Atlanta, GA 30314, USA}
\address{${}^{\dag}$Centro per lo studio dei sistemi dinamici, Universit\'a\\
degli studi dell'Insubria, 22100 Como, Italy}
\date{\today}
\maketitle

\begin{abstract}
We discuss the construction of optimized electronic filters using inverse
scattering methods. We study a wide range of densities and temperatures, 
{\it room temperature included}. Discretization methods of the potential (%
{\it including the self-consistent potential of the conduction electrons})
are worked out that retain all its properties.
\end{abstract}

\pacs{85.30.Dc, 03.65.Nk, 73.20.D}

\narrowtext

\section{Introduction}

A semiconductor heterostructure can be modeled by a system of equations
describing (with a certain degree of completeness and precision) the state
of the system. The equations depend on a set of structural and compositional
data: the spatial dependence of the chemical composition (including dopant
profiles), the applied external fields, etc. The system behavior (response)
is described by functional data, such as the electric or thermal
conductance, the energy dependence of the electron transmittance, the
wavelength dependence of the optical absorption coefficient, etc. The
functional data can be computed using the solution of the equations, and are
thus functionals of the structural and compositional data.

To design a heterostructure for a certain application is to find a set of
structural and compositional data, which is physically (and technologically)
achievable, such that the values of a chosen subset of functional data will
be within a desirable range. The designer, in principle, solves an inverse
problem: inverting the dependence of the functional data on the structural
and compositional data.

In the usual approach, the starting point is a proposed configuration of the
structural and compositional data which is then varied to optimize the
component's performance; at each iteration the spectral-scattering data
describing the performance have to be re-calculated. In this method, much
depends on the initial choice of parameters as in general several non
equivalent local extrema of the response are present in the parameter space.

Inverse methods, when available, are not subject to this drawback. Contrary
to the previous situation, the designer starts from data describing the
system's performance, such as current spectral response curves and
determines the optimal structural and compositional data. He automatically
gets an optimized set of parameters. Even if the model used to make the
optimization calculation feasible lacks all the details of the full physical
situation, one has still the possibility to use the traditional approach to
fine tune the results from such a new starting point.

The inverse method we use requires three different steps \cite{eur,micr}:

1. The reconstruction of the phase of the transmittivity of a quantum device
using Finite Dispersion Relations under the very convenient form of the Pad%
\'{e} approximations.

2. The Standard Inverse Scattering method, making use of the results of Kay
and Sabatier \cite{kay,tier} for the inverse problem with rational
coefficients (Pad\'{e} approximations). This allows to build an efficient
algorithm for solving the inverse problem and obtaining the potential.

3. A unitary transformation that maps the place dependent mass problem of a
BenDaniel equation onto the usual constant mass Schroedinger problem keeping
the potential in the equation {\it local} [2].

By applying the previous procedure, it becomes possible to design new
realistic quantum wells \cite{lev,cap} submitted to {\it optimization
constraints}. In particular, one can use inverse methods to design and
fabricate novel color sensitive QWIPs in the medium- and long-wave IR.
Moreover, one of the most unexpected and astonishing results of the inverse
method is that inclusion of the self-consistent potential of the conduction
electrons is far much simpler than in the usual direct approach \cite{micr}.

As a first application, we will use the versatility of the inverse
scattering methods for the Schroedinger/BenDaniel-Poisson model \cite%
{eur,micr}, for designing and building prototypes of improved electronic
filters.

An aspect that has not yet been dealt with in depth is how to discretize the
potential profiles obtained by inverse scattering methods so as to make
their production viable while mantaining at the same time their properties,
their electron wave reflectance in particular. An added complication is the
necessary presence in the devices of substantial densities of conduction
electrons: the electric potentials they generate do not, if suitably
compensated by doping the devices \cite{micr}, significantly alter the
reflectance of the "smooth" potentials obtained by inverse scattering; but
what about the discretized potentials? The aim of the present paper is to
deal with the problems stemming from this necessity to both discretize the
potentials and have substantial densities of conduction electrons. To avoid
unnecessary complications, we shall skip the last step of the procedure
described in Ref. \cite{micr}, namely the mapping of the (constant mass)
Schroedinger equation into the (variable mass) BenDaniel and Duke equation,
as it has no direct bearing on the present discussion.

Throughout the present paper we shall make use of the "natural" units thus
defined: energy (e.u.): $0.1$ $eV$; mass: $0.1$ $m_e$; length: $1$ $nm$;
time: $2.3845$ $fs$; in these units $\hbar=2.7604$.

\section{Inverse Scattering Problem}

The procedure to build a potential having a given reflectance profile for
elecron waves and no bound state is described in detail in Ref. \cite{micr};
here we wish to stress that while standard type II Pad\'{e} \cite{pad}
fitting routines can be a useful first step in building a Pad\'{e}
approximant, it is often the case that the parameters thus found must be
adjusted, as care must be taken to avoid obtaining a potential either
impossible to build because exceeding in height the available potential
difference (for $Al_{c}Ga_{1-c}As$, where $c$ is the Aluminum concentration,
this value is about $3e.u.$) or impractical to grow because having a support
too wide and/or too many peaks. The reconstruction of the potential makes
use of the poles and relative residues of the reflection coefficient $%
R_{+}(k)$ for plane waves incoming from $+\infty $ where $k=\sqrt{mE/\hbar
^{2}}$ is the electron momentum. The frequencies of oscillation relate to
the real part ($a_{j}$) of the poles (where $j=1,n$, $n$ being the number of
poles), the decay factors to the imaginary part ($b_{j}$), and the
amplitudes to the residues ($\rho _{j}$). Large values of the $a_{j}$'s give
fast oscillations of the potential; small values of the $b_{j}$'s result in
slow decays of the oscillations and therefore in a wide support for the
potential; finally, large $\rho _{j}$'s -that are often the case when poles
are very close to each other- result in large oscillations of the potential;%
{\it \ all three must therefore be avoided.} Sometimes these conditions can
be satisfied with no problems, sometimes instead they force us to some kind
of compromise. An example of the latter case is a bandpass filter: since the
more the poles are close to the real axis and close to each other the more
the transmission plateau is flat and close to unity, some extra care must be
taken when choosing the Pad\'{e} approximant parameters.

\section{Discretization of the Potential}

A continuous potential as the one obtained by inverse scattering \cite{micr}
is impractical to build; we must therefore discretize it by a sequence as
small as possible of flat steps. Several options are possible, the choice
depending on the characteristics of the heterostructure we wish to build:

{\it a)} placing the steps at the inflection points of the original
potential and choosing their height to be equal to that of the original
potential gives us an (almost) zero transmittance in the forbidden zones and
an almost satisfactory plateau in Fig. \ref{fig2} without significant
shifts. A desirable characteristic of this method is {\it the unexpected
reduced width of the two peaks in Fig. \ref{fig1}.} A second scheme can be
used to enhance this characteristic:

{\it b)} If we move the position of the steps to the zeroes of the original
potential, both peaks are now shifted, but we have succeeded {\it to reduce
the width of both of them by a factor of }$4$!

As a rule, both methods should be tested and the one that gives the result
nearest to the desired one retained.

\section{Doping for Non-zero Density of Conduction Electrons}

The (selfconsistent) electric field generated by the conduction electrons
can seriously modify the effective potential and thus alter the transmission
profile. It must therefore be compensated. We achieve this in two steps:
first, from the potential obtained by inverse scattering we subtract the
selfconsistent field of the electrons themselves and of some suitable doping
charge such that the total charge and dipole moment in the device are zero
(see Ref. \cite{micr}); then, after obtaining the "chemical composition
potential" by discretization of the resulting potential, we calculate the
selfconsistent field for this new potential itself, adjusting the position
and value of the doping charge, so that again the total charge and dipole in
the device are zero and the effective potential is as close as possible to
the "chemical composition potential" itself. An iterative selfconsistent
calculation starting from the doping charge and dipole moment obtained for
the potential before discretization converges{\it \ in the worst case we
considered in less than ten steps to a precision of five digits. }Note that,
once the correct doping has been added, the changes to the potential are
usually very small; on the other hand a wrong doping is bound to seriously
deteriorate the transmittance profile. In both of the following examples we
assumed the doping to be in the form of a sheet charge; this
approximation is valid as long as the doping can be kept in a region more
narrow than the steps of the potential; should this not be the case, a
suitable charge width should be considered and taken into account, requiring
only minor changes to the codes.

\section{Two Examples}

To illustrate the above procedure, we now give two examples. The starting
point for both is a Pad\'e approximant to the reflectance of the form: 
\begin{eqnarray}
{\cal R}(E)={\frac{{{\cal A}_{2N}(E)}}{{({\cal A}_{2N}(E)+E{\cal K}%
_{2(N+2)}(E))}}}  \label{pad} \\
{\cal K}_{2(N+2)}(E) = \Pi_{j=1,N+2}\left[\left(E-F_j\right)^2+\delta_j^2%
\right] \hspace{0.5in}  \label{zeT} \\
{\cal A}_{2N}(E) = g\cdot\left[\Pi_{i=1,N}\left(E-E_i\right)\right]^2 
\hspace{0.15in}E_i,F_j,\delta_j \in {\bf R}^+  \label{zeR}
\end{eqnarray}
Eq. (\ref{pad}) guarantees that ${\cal R}(0) = 1$; it was already given in
Ref. \cite{micr} but there higher degrees for ${\cal K}_{2(N+2)}(E)$ were
allowed. We choose to use the lowest possible degree ensuring the absence of 
$\delta$-function terms in the potential \cite{sab}. Eq. (\ref{zeT})
guarantees that, excepted the zero at the origin, the transmittance has
complex conjugate zeroes; Eq. (\ref{zeR}) instead guarantees that the
reflectance has real double zeroes. The three equations together guarantee
that $0 \leq {\cal R}(E) < 1$ for $E > 0$.

We assume the devices to be grown from $Al_cGa_{1-c}As$; we therefore use as
parameters those for the substrate $GaAs$: effective electron mass $%
m_{\infty}=0.67$ in natural units, and electrical permeability $%
\varepsilon=13$.

\subsection{A Filter with Two Narrow Transmission Resonances}

Our construction of a filter with two narrow peaks,{\it \ the one at higher
energy being the narrower }\cite{priv}, starts from a Pad\'{e} approximant having
the same number of poles and zeroes as the one given in Refs. \cite{micr,eur}
but in the form (\ref{pad}) given above with: $N=2$, $g=1.5$, $E_{1}=0.4e.u.$%
, $E_{2}=1.1e.u.$, $F_{1}=0.2e.u.$, $\delta _{1}=0.05e.u.$, $F_{2}=0.8e.u.$, 
$\delta _{2}=0.01e.u.$, $F_{3}=1.2e.u.$, $\delta _{3}=0.10e.u.$, $%
F_{4}=1.5e.u.$, and $\delta _{4}=0.20e.u..$ $E_{1}$ and $E_{2}$ give the
positions of the transmittance peaks; the four $F_{j}$ frame them and
determine the peaks' width.

The final transmittance using procedure {\it b)} to discretize the
potential, is shown in Fig. \ref{fig1} for three temperatures ($T=70$, $233$%
, and $300$ $K$) and two background electron densities ($10^{13}$, $10^{15}$%
); the six curves are undistinguishable. The compensating doping charges and
their positions are given in table $1$. We have also studied other
temperatures and densities that are not reported for lack of space. Only at
the highest density studied ($10^{17}$ $el/cm^{3}$) the reflectance is
strongly affected by the electron density, especially at low temperatures.
The reason is that since at low temperatures the deviation from the
background value of the electron density is less spread out, it has sharper
peaks which are reflected in analogous sharp peaks in the selfconsistent
electrical field; the plateaus of the discretized potential are therefore
more strongly deformed than at higher temperatures.

\subsection{A Bandpass Filter}

For this case some preliminary remarks are in order: the reflectance of an
ideal bandpass filter should be zero inside the band and one everywhere
else. On the other hand for a (piecewise) continuous potential that decays
at infinity we have:

{\it a)} ${\cal R}(E)\rightarrow 0$ for $E\rightarrow \infty$.

As a consequence, the forbidden region will have to be finite; it will be
our aim to make it as wide as possible on both sides of the transmission
band.

{\it b)} ${\cal R}(E)=1$ only for $E=0$; otherwise ${\cal R}(E) < 1$.

Therefore, there will be a residual transmittance in the forbidden region;
it will be our aim to make it as small as possible.

{\it c)} ${\cal R}(E)$ can be zero only on a set of points which is at most
countable.

Because of this, the desired transmission band will have to be built by a
superposition of peaks, close enough that the transmission plateau is almost
flat and sharp enough that the transition from forbidden region to allowed
band is narrow compared to the band width.

We found convenient to choose the following parameters in eq. (\ref{pad}): $%
N=3$, $g = 20$, $E_1 = 0.9 e.u.$, $E_2 = 1.1 e.u.$, $E_3 = 1.3 e.u.$, $F_1 =
F_2 = 0.3 e.u.$, $\delta_1 = 0.08 e.u.$, $\delta_2 = 0.22 e.u.$, $F_3 = F_4
= F_5 = 1.6 e.u.$, $\delta_3 = 0.48 e.u.$, $\delta_4 = 0.30 e.u.$, and $%
\delta_5 = 0.16 e.u.$. The transmittance plateau is given by the three
reflectance zeroes in $E_i$, $i=1,2,3$; the five $F_j$ frame them and
determine the two forbidden zones at each side of the allowed band.

The final transmittance using procedure {\it a)} to discretize the
potential, shown in Fig. \ref{fig2} for the same three temperatures and two
background electron densities used in the previous example, is again equal
in all cases; the compensating doping charges and their positions are given
in table $2$. Again we found that only at the highest density studied ($%
10^{17}$ $el/cm^{3}$) the reflectance is strongly affected by the electron
density.

\section{Conclusions and Aknowledgements}

We have shown that both discretization of the potential and substantial
densities of conduction electrons can be dealt with in ways that preserve
the essential characters of the desired reflectance. We should note that in
both the examples given the background electron distribution is repelled
from the potential and therefore the required compensating doping is of the
N-type; this can sometimes be impractical. A way to circumvent the problem
is to add a (monodimensional) bound state via a Darboux transformation \cite%
{sab}: the high electron density in it ensures that the doping be of the
P-type.

Finally, we would be glad to receive feed-back from readers. We are ready to
analyze and design electronic filters that are difficult or impossible to
deal with by the ordinary method.

We wish to aknowledge the participation to the first phases of the present research of the late G.A. Mezincescu, who has been for a long time a driving force in our group.

\begin{figure}[htbp]
\caption{The reflectance of our double peak filter at $T=70, 233,$ and $300$ 
$K$ for two different densities of conduction electrons ($10^{13}$ $el/cm^3$
and at $10^{15}$ $el/cm^3$): all six curves are indistinguishable on this
scale from the "optimal" reflectance obtained using procedure {\it a)}
described in the text and truncating the support at $50$ $nm$. The width ratio of the two peaks is approximately $3$.}
\label{fig1}
\end{figure}

\begin{figure}[htbp]
\caption{The reflectance of our bandpass filter at $T=70, 233,$ and $300$ $K$
for two different densities of conduction electrons ($10^{13}$ $el/cm^3$ and
at $10^{15}$ $el/cm^3$): all six curves are indistinguishable on this scale
from the "optimal" reflectance obtained using procedure {\it b)} described
in the text and truncating the support at $100$ $nm$.}
\label{fig2}
\end{figure}

\begin{table}[tbp]
\begin{eqnarray}
\hline\hline  \nonumber \\
T(K)\hspace{0.05in} el. density (el/cm^3)\hspace{0.05in} charge(cm^{-2})%
\hspace{0.05in} position (nm)  \nonumber \\
\hline  \nonumber \\
70 \hspace{0.7in} 10^{13} \hspace{0.3in} -2.3239\ast 10^{7} \hspace{0.65in}
9.0  \nonumber \\
70 \hspace{0.7in} 10^{15} \hspace{0.3in} -2.3008\ast 10^{9} \hspace{0.65in}
9.0  \nonumber \\
233 \hspace{0.7in} 10^{13} \hspace{0.3in} -1.8134\ast 10^{7} \hspace{0.65in}
6.8  \nonumber \\
233 \hspace{0.7in} 10^{15} \hspace{0.3in} -1.8113\ast 10^{9} \hspace{0.65in}
6.8  \nonumber \\
300 \hspace{0.7in} 10^{13} \hspace{0.3in} -1.6038\ast 10^{7} \hspace{0.65in}
6.5  \nonumber \\
300 \hspace{0.7in} 10^{15} \hspace{0.3in} -1.6061\ast 10^{9} \hspace{0.65in}
6.6  \nonumber \\
\hline \hline \nonumber
\end{eqnarray}%
\caption{Doping charge and position at different temperatures and background
electron densities for the filter with two narrow transmission peaks
described in the text.}
\end{table}

\begin{table}[tbp]
\begin{eqnarray}
\hline\hline  \nonumber \\
T(K)\hspace{0.05in} el. density (el/cm^3)\hspace{0.05in} charge(cm^{-2})%
\hspace{0.05in} position (nm)  \nonumber \\
\hline  \nonumber \\
70 \hspace{0.7in} 10^{13} \hspace{0.3in} -3.1729\ast 10^{7} \hspace{0.6in}
12.8  \nonumber \\
70 \hspace{0.7in} 10^{15} \hspace{0.3in} -3.1455\ast 10^{9} \hspace{0.6in}
12.9  \nonumber \\
233 \hspace{0.7in} 10^{13} \hspace{0.3in} -1.8134\ast 10^{7} \hspace{0.6in}
10.4  \nonumber \\
233 \hspace{0.7in} 10^{15} \hspace{0.3in} -1.8113\ast 10^{9} \hspace{0.6in}
10.3  \nonumber \\
300 \hspace{0.7in} 10^{13} \hspace{0.3in} -1.6038\ast 10^{7} \hspace{0.6in}
10.1  \nonumber \\
300 \hspace{0.7in} 10^{15} \hspace{0.3in} -1.6061\ast 10^{9} \hspace{0.6in}
10.1  \nonumber \\
\hline \hline \nonumber
\end{eqnarray}%
\caption{Doping charge and position at different temperatures and background
electron densities for the bandpass filter described in the text.}
\end{table}

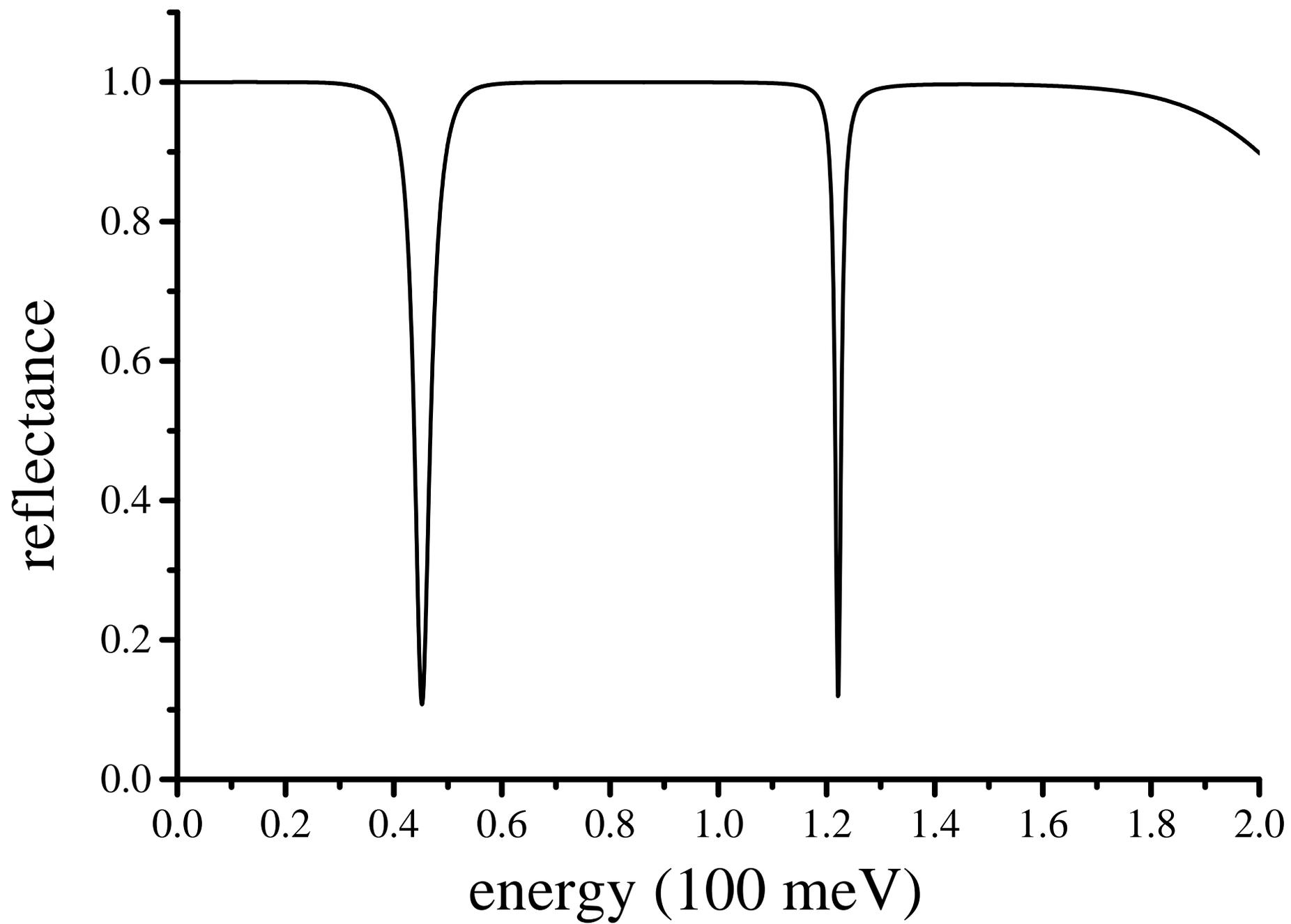
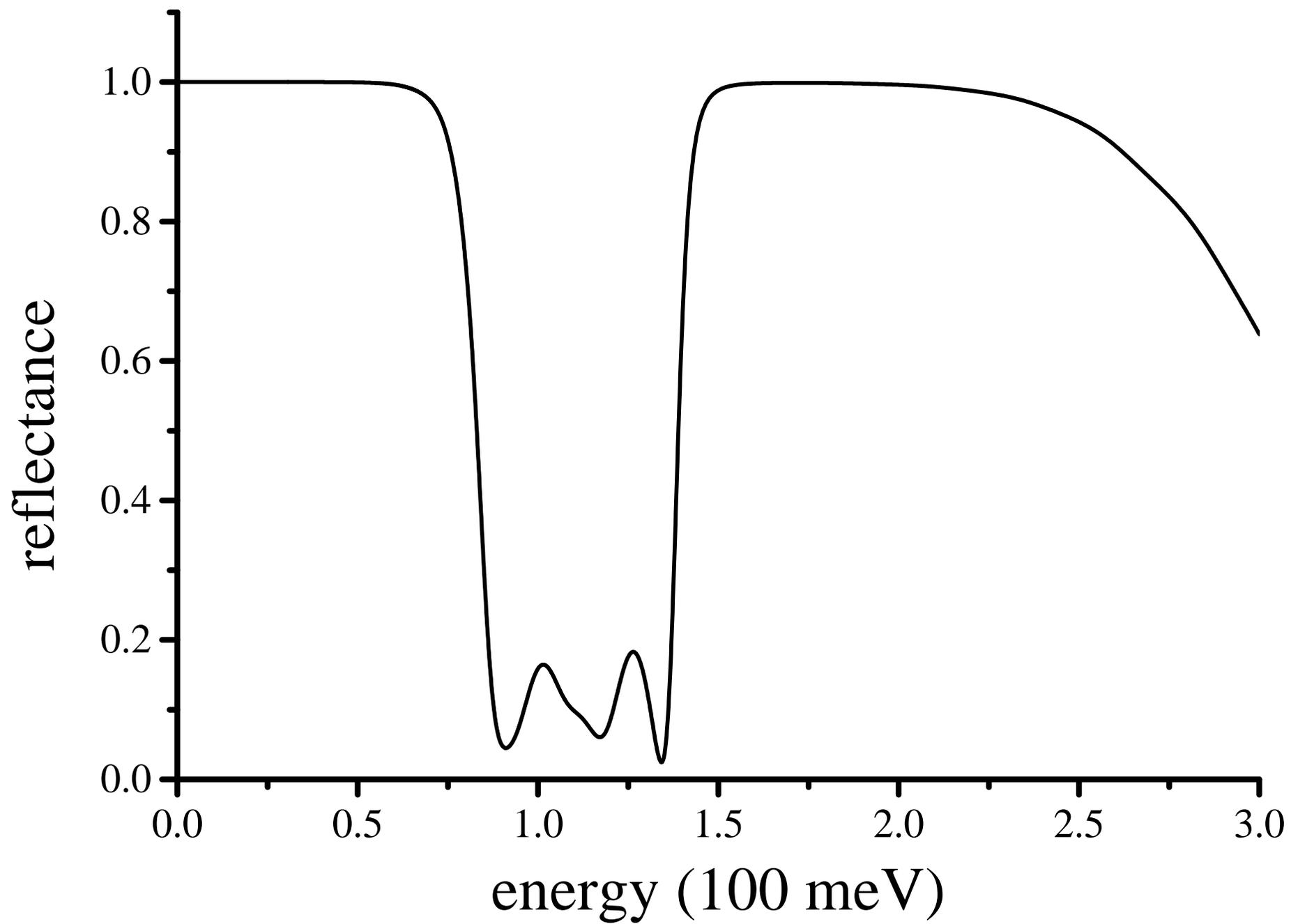
\end{document}